\documentclass[prl,aps,twocolumn,floats]{revtex4}
\usepackage{amsmath}
\usepackage{graphicx}
\usepackage{epsfig}

\newcommand{\be}{\begin{equation}}
\newcommand{\ee}{\end{equation}}
\newcommand{\bea}{\begin{eqnarray}}
\newcommand{\eea}{\end{eqnarray}}

\begin{document}
\title{Scaling theory of magneto-resistance in disordered local moment ferromagnets}

\author{Gergely Zar\'and$^{1}$, C\u at\u alin Pa\c scu  Moca$^{2}$
and Boldizs\'ar Jank\'o$^{3,4}$}

\affiliation{ $^1$ Budapest University of Technology and Economics,
H-1521 Budapest, Hungary
\\
$^2$Department of Physics, University of Oradea, 410087 Oradea,
Romania
\\
$^3$Department of Physics and Institute for Theoretical Sciences,
University of Notre Dame, Notre Dame, Indiana, 46556
\\
$^4$Materials Science Division, Argonne National Laboratory, 9700
South Cass Avenue, Argonne, Illinois 60439
\\
}
\date{\today}

\begin{abstract}
We present a scaling theory of magneto-transport in
Anderson-localized disordered ferromagnets. Within our framework a
pronounced magnetic-field-sensitive resistance peak emerges
naturally for temperatures near the magnetic phase transition. We find that
the resistance anomaly is a direct consequence of the change in
localization length caused by the magnetic transition. For
increasing values of the external magnetic field, the resistance
peak is gradually depleted and pushed towards higher temperatures.
Our results are in good agreement with magneto-resistance
measurements on a variety of disordered magnets.
\end{abstract}

\maketitle

Complex magnetic materials showing strong magneto-resistance have
simultaneously been at the focus of the attention of the magnetic
recording industry and the field of strongly correlated electron
systems. As a consequence, the interplay between electronic and
magnetic degrees of freedom has been intensely studied both
experimentally and theoretically during the last four decades.

One of the first theoretical frameworks to provide guidance for
understanding magneto-resistance measurements on magnetic metals were
provided by de Gennes and Friedel \cite{degennes} who predicted that
long range magnetic fluctuations near the Curie temperature $T_C$
will result in a cusp-like singularity of the resistivity $\rho (T)$.
As subsequent experiments failed to observe such
singular behavior, Fisher and Langer \cite{langfish} pointed out
that $\rho (T)$ shows no anomaly, but it is $d \rho (T)/dT$ that
bears resemblance to a cusp near $T_c$, and it is caused by short
(not long) range magnetic fluctuations.

While the work of Fisher and Langer provided for several decades a
very useful theoretical framework to understand magneto-resistance in
itinerant magnets, recently discovered complex magnetic materials,
such as manganites \cite{manganites}, and diluted magnetic semiconductors \cite{dms}
{\it do show} a resistivity peak directly in  $\rho (T)$  and a corresponding anomaly 
in a relatively large temperature window near  $T_c$. For magnetic semiconductors and for some of the manganites
the before-mentioned  resistivity peak  appears to be located precisely at $T_c$.
As pointed out by Littlewood and his coworkers \cite{littlewood}, such behavior,
especially for non-metallic samples, seems to fall outside the range of applicability of the
Fisher-Langer model. Several proposals have been made, involving magnetic polarons \cite{littlewood}, critical
spin-flip scattering \cite{ohno}, thermal magnetic fluctuations within the six-band model \cite{brey}
or Nagaev's magneto-impurity scattering model \cite{nagaev}. Most of these theories focus
on metallic samples and on producing the resistive anomaly near $T_C$, while ignoring non-metallic samples and
at least part of the full temperature range.

The available experimental results, however, provide  severe constraints if one
 requires  the theoretical framework to reproduce not only the  relative shape and position
of the anomaly in a narrow temperature window near $T_C$ (fluctuation regime), but the actual {\it magnitude} of
the anomaly, together with the resistive behavior in the paramagnetic {\it and}
the magnetically ordered phase as well. Furthermore, 
the theory should explain the fact that,  with the application of an external magnetic field,
the experimentally observed anomaly is simultaneously reduced in height and pushed towards higher temperatures
and the magneto-resistance curves for different external fields never cross. A successful
theoretical framework for the magneto-resistance anomaly in disordered magnets should simultaneously satisfy all
the experimental constraints mentioned above.

In this paper we show how all properties  described above
can be explained  as a consequence of the interplay between disorder induced  localization
transition  and magnetism, without invoking additional mechanisms such as Jahn-Teller 
distortion, probably important for some manganites \cite{Millis}.  
Our theory is based on simple scaling argument and is
relevant for a large number of systems including some of the  manganites and a number 
of magnetic semiconductors. 

Let us first discuss how the original one-parameter scaling theory of localization
has to be modified to describe magnetic systems. In the standard scaling theory of
localization one argues that the typical dimensionless conductance $g(2L)$ of a cube of linear size size $2L$
is uniquely determined by the typical dimensionless conductance $g(L)$ of its pieces of size $L$. This assumption is
summarized in the following scaling equation:
\begin{equation}
\frac{d\ln g}{d\ln L} =\beta(g)\;.
\label{eq:beta}
\end{equation}
The $\beta (g)$ function above depends exclusively on $g$, the dimension, and the {\em symmetry} of the Hamiltonian,
however, it does not depend on other microscopic properties of the disorder. While $\beta(g)$ can be evaluated
for large values of $g$ perturbatively, in order to determine its full shape, numerical computations are needed.
For very large $g$, $\beta \approx 1$ in three dimensions, while for very small $g$ it scales
as $\beta(g) \approx \ln g$ \cite{gangof4,localization_review}.

The scaling equation above can be used as a powerful tool to analyze the localization transition.
Very importantly, in three dimensions the $\beta$ function always vanishes at a critical value
$g_c$ of $g$, $\beta(g_c) = 0$.  This zero of the $\beta$ function is associated with the localization transition:
Consider a small piece of mesoscopic size $l_0$ having a microscopic conductance  $g_0$. If $g_0> g_c$, then the zero temperature
dimensionless  conductance  increases as we increase $L$ and asymptotically behaves as $g(L) \sim \sigma L^{d-1} $, with $\sigma(g_0)$
the conductivity.   For $g<g_c$, on the other hand, the conductance scales as $g(L) \sim \exp(- 2L/\xi)$, 
with $\xi(g_0)$ the localization length. From (\ref{eq:beta}) it follows that  for
$g_0 \to g_c$ the localization length diverges on the localized side as
$\xi \sim (g_c - g_0) ^{-\nu}$ while the conductivity goes to 0 on the metallic side of the transition as
$\sigma \sim (g_0-g_c)^\nu$ in three dimensions.  The critical exponent $\nu$ is related to the slope of the beta function at $g_c$ as
$1/\nu \equiv d \beta/d{\ln}g|_{g_c}$.

Let us now consider a disordered local-moment ferromagnet, in which local moments ${\vec \Omega}_i$ coupled to some
charge carriers are responsible for the magnetism. The first observation we make is that - compared to electronic
processes - spin fluctuations are usually slow. This is especially true in the vicinity of the ferromagnetic phase
transition where the conductance peak of our interest appears.
This implies that for transport properties the magnetic moments can be treated
as static scatterers, and can be replaced by classical spins $\vec \Omega_i$. While scattering from these static
magnetic moments  itself is typically  not sufficient to lead to localized charge carriers, it can substantially
increase the effect of static disorder, and help to localize charge carriers.

The $T=0$ conductance of a sample of size $L$ will thus depend on the particular distribution of magnetic moments
$P(\{\vec \Omega_i\})$ \cite{note}.  One can argue, however, that the
$\beta$ function (\ref{eq:beta}) should depend on this distribution only through the conductance 
$g_0$. Thus the effect of magnetic moments
appears in two ways (a) It determines the appropriate symmetry class of the $\beta$ function (Unitary Ensemble =
UE), and (b) enters the {\em microscopic conductance} $g_0 = g_0(P\{\omega\}) = g_0[P]$.
In general, $g_0$ is therefore a complicated  non-universal function of temperature,
and magnetic field, $g_0= g_0(T/T_C, H/T_C)$ which also depends on the
microscopic details of the system. However, once one knows this function and the $\beta$-function,
one can use the scaling theory to determine transport properties of the system, 
as we shall discuss below. One possibility to determine $g_0$ is to perform, {\em e.g.}, 
a Monte Carlo simulation for a small system, and use the computed conductivity as 
$g_0$. 

The scaling equation provides us the conductivity (or localization length) of the electrons,
provided that the electronic temperature is zero. In a real system, however, finite 
temperature has a dual role: On one hand, it changes the distribution $P$ and thus the 
value of $g_0(P)$, but one must also take it into account
through the temperature of the conduction electrons. This is a rather complicated task on the 
metallic side, where the temperature-dependent dephasing length  $L_{\rm max} = L_\varphi(T)$   
provides a cut-off for the scaling,
with $L_\varphi(T)$ being a non-universal function of the temperature $T$. It is, however,
 simple to incorporate the effects of finite electronic temperature on the insulating side
via  Mott's variable range  hopping formula \cite{Mott}:
\begin{equation}
\sigma  \sim {\rm exp}\bigl \{ -C \bigl(\frac {\Delta_\xi(P)}{T}\bigr)^{1/4} \bigr\}\;,
\label{eq:Mott}
\end{equation}
where $\Delta_\xi(P) = 1/N_0 \xi^d(P)$, with $N_0$ the density of states at the Fermi level, and $C$ is a constant of the 
order of unity \cite{foot_note_gap}.  Note that in this formula
$\xi$ must be determined from the integration of the scaling equation, and it therefore depends on
$g_0(P)$ and thus on the temperature through the distribution $P$.

In order to illustrate the ideas described above, let us consider the disordered
Kondo lattice with classical spins:
\begin{eqnarray}
H_K & = &-t \sum_{(i,j),\alpha} c^\dagger_{i\alpha} c_{j\alpha} + \sum_{i,\alpha} \epsilon_i
c^\dagger_{i\alpha} c_{i\alpha} \nonumber \\
&+& J \sum_{i,\alpha,\alpha'} {\vec \Omega}_i c^\dagger_{i\alpha}{\vec \sigma}_{\alpha\alpha'} c_{i\alpha'}
\;.
\label{Kondo}
\end{eqnarray}
Here  $t\equiv1 $ denotes the hopping amplitude of the conduction electrons on a lattice,
$c^\dagger_{i\alpha}$ creates an electron with spin $\alpha$ at site $i$, and $J$ is the 
exchange coupling between the spin of the electrons and the classical local 
moments $\vec \Omega_i$. The $\epsilon_i$-s in
Eq.~(\ref{Kondo}) denote random on-site energies, which we generate with a uniform distribution
between  $\pm W$. In what follows, we shall concentrate on the $J\gg W,t$ limit of
 Eq.~(\ref{Kondo}), which is relevant for strongly spin-polarized systems such as  
manganites \cite{Millis} and some ferromagnetic semiconductors with polarized 
impurity bands \cite{polarized_ferromagn_semic}. In this limit Eq.~(\ref{Kondo}) simplifies to:
\begin{equation}
H_{J=\infty}  = - \sum_{(i,j)} t_{ij} a^\dagger_{i} a_{j} + \sum_{i} \epsilon_i
a^\dagger_{i} a_{i}\;,
\label{projected}
\end{equation}
where the $a_i$ denote spinless Fermions corresponding to the original Fermions aligned 
antiparallel with the local moments, and $t_{ij} = e^{-i\varphi_{ij}} (1 + {\vec \Omega_i} 
{\vec \Omega_j})/2$, with $\varphi_{ij}$ a Berry phase that depends on the 
directions $\vec \Omega_i$ and $\vec \Omega_j$ \cite{AndersonHasegawa}.
The electronic properties of Eq.~(\ref{projected}) have been analyzed for completely 
radom spin orientations in Ref.~\cite{Bishop}. Here, however, we want to 
study the effect  of the ordering of the spins on the electronic properties.

In principle, the finite temperature distribution function of the spins could be 
computed using an effective  action that one obtains by integrating out the conduction 
electrons at temperature $T$. 
Instead of doing this, we shall replace this effective action by a simple mean field theory, 
and assume that the distribution of $\vec \Omega_i$ is simply
given by
\begin{equation}
P_0(\vec \Omega_i) \equiv \frac1Z  {\rm exp}[ \Omega_i^z \alpha]\;,
\label{P0}
\end{equation}
where $\alpha = (H + K \langle \Omega^z\rangle)/T]$. Here
$H$ denotes an external field, $K$ is an effective exchange coupling between the spins, and
the magnetization $m=\langle \Omega^z\rangle$ must be determined self-consistently.
It is not difficult to see that in this case
$P(\vec \Omega_i)$ depends in fact on a single parameter  $m\equiv \langle  \Omega^z \rangle$,
which is a universal  function of $t=T/T_C$ and $h = H/T_C$, with $T_C = K/3$ being 
the critical temperature. We computed $m(t,h)$ numerically by solving 
simple transcendental equations.

\begin{figure}[tb]
\begin{center}
\epsfig{figure=phase_diagram_gergely.eps, width=6.5 cm,clip}
\\
\epsfig{figure=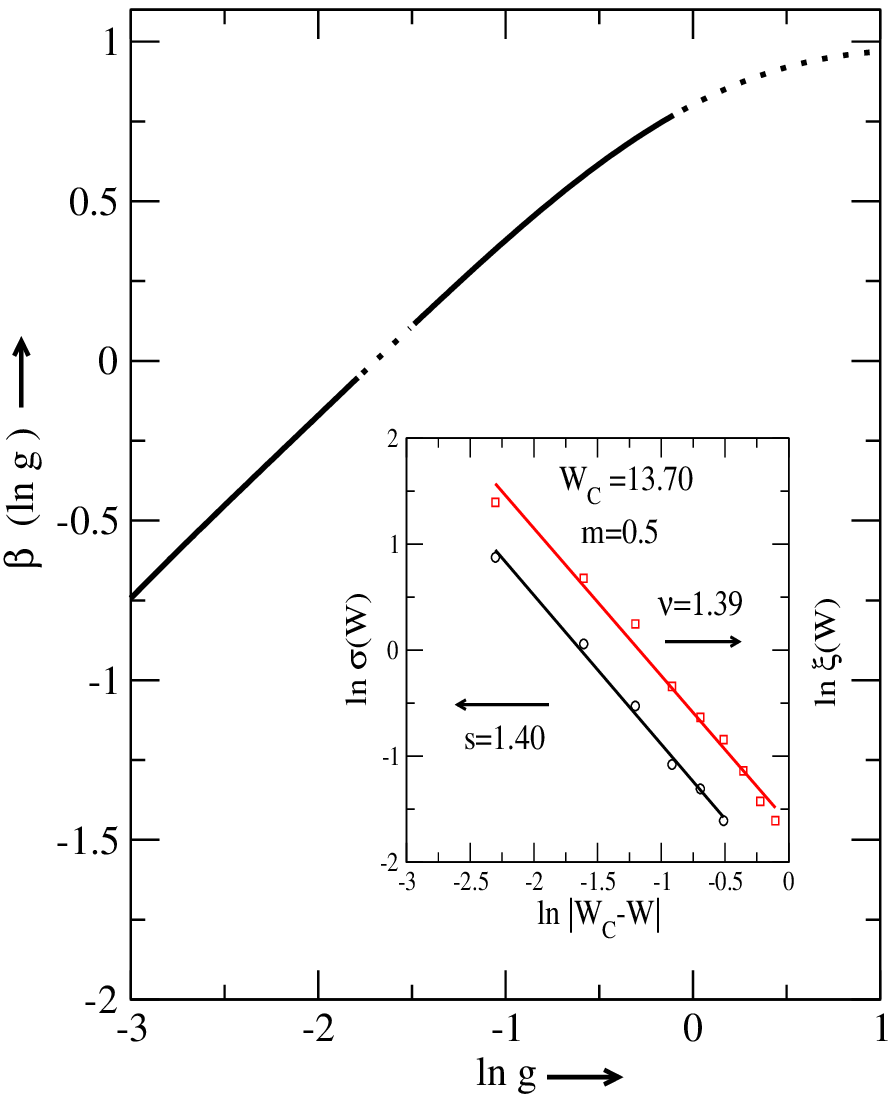, width=6.5cm,clip}
\end{center}
\vskip0.1cm
\caption{\label{fig:beta}
Top: Zero temperature localization phase diagram of the Kondo lattice model 
defined in the text obtained from  a finite size scaling analysis of Ljapunov 
exponents (dots). The localization phase transition is shifted towards stronger disorder as 
we align the spins. The dashed line denotes the phase boundary obtained using the naive estimate,
Eq.~(\ref{g_0(m)}). Bottom: Beta function obtained numerically from the scaling analysis. The inset shows
the divergence of the  localization length and the resistivity at the critical disorder for 
the case $m=0$. We estimate the critical exponent to be $\nu\approx 1.4$.}
\end{figure}

To obtain the phase diagram of the model defined by Eqs.~(\ref{projected}) and (\ref{P0}) we carried out
a careful transfer  matrix  analysis within the formalism following the original
work of McKinnon and Kramer \cite{KramersMcKinnon,Pascu}. The main results of our analysis are summarized
in Fig.~\ref{fig:beta} for a carrier concentration 0.5 electron/spin.
Firstly, we find that larger and larger disorder is needed to obtain a localized
phase for increasing $m$ (see the phase boundary in Fig.\ref{fig:beta}a).
In other words, aligning spins delocalizes electrons, and leads to a decrease of the resistivity.
We performed a scaling analysis of the Ljapunov exponents and found that for $m<0.9$ 
all data collapsed to a single scaling curve, independent of the specific value of $m$, 
confirming the single parameter scaling  hypothesis made earlier. 
Note, however, that the  $m\approx 1$ data could not be collapsed with the $m<1$ data.
This is quite natural, since for $m=1$ the Hamiltonian belongs to  a different symmetry class
(Orthogonal Ensemble).  The scaling analysis also made us possible to estimate 
the $\beta$ function shown in Fig. \ref{fig:beta}b. The critical exponent $\nu \approx 1.4$ 
is in good agreement with the earlier results of Ref.~\cite{Ohtsuki} for the unitary ensamble.

The phase boundary can be qualitatively understood if we assume that the microscopic
 conductance is proportional to the conductance of a single bond. 
 In the most naive 
approximations, the effect of the magnetic field is just to reduce the effective value 
of the hopping, $t^2\to t_{\rm eff}^2 = t^2 f(m)$, and the conductance is  
roughly proportional to $\sim t_{\rm eff}^2/W^2$,
\begin{equation}
g_0(m) \approx C {t_{\rm eff}^2\over W^2} = {\tilde g}_0 \;f(m)\;,
\label{g_0(m)}
\end{equation}
where $\tilde g_0 = g_0(m=1)$ denotes the microscopic conductance for fully aligned spins.
The function $f(m)=f(m(h,t))$ can be obtained in this approximation from the phase boundary, 
which is determined by the condition $g_0(m) = g_c$, and is simply given
by $W_c(m) = W_c(m=1) \sqrt{f(m)}$. As shown in Fig.~\ref{fig:beta}a, the simple function
$f(m) = (1 + m^2)/2$ gives a very reasonable agreement with the numerical phase boundary. 

Having $f(m)$ and the $\beta$-function at hand, we can now carry out the program outlined 
above and combine Eqs.~(\ref{eq:beta}), (\ref{eq:Mott}), and  (\ref{g_0(m)}) to determine the
temperature-dependence of the conductivity in the localized phase.
For the sake of simplicity, we shall assume  that  we are close enough to the metal-insulator 
transition, and approximate the $\beta$-function as $\beta(g) \approx \frac1\nu \ln (g/g_c)$ 
with $\nu \approx 1.4$. Since $\nu$ is not far from  1, this is a reasonable approximation. 
In this case the resistivity can be expressed as
\begin{equation}
{\rm ln} \varrho = A \left[ {\rm ln} \frac{g_c}{g_0(m)} \right]^{3\nu/4} \frac1{t^{1/4}} ;,
\label{eq:logro}
\end{equation}
where $A$ is a constant of the order of unity, and  $g_c / g_0(m) = B/f(m(t,h))$. The 
constant $B$ here measures simply the  distance from the localized phase, $B = g_c/g_0(m=1)$, and
$f(m)$ is the scaling function in Eq.~(\ref{g_0(m)}).

\begin{figure}[tb]
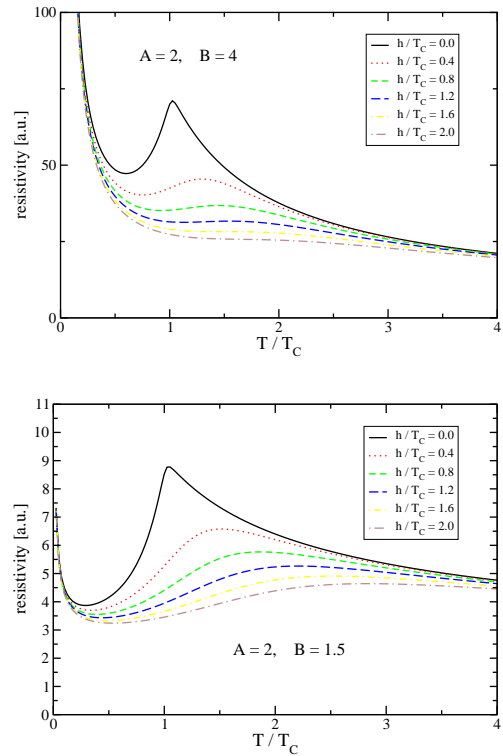

\begin{center}
\epsfig{figure=resist_b4_a2_nu1_4.eps, width=6.5 cm,clip}
\vskip0.5cm
\epsfig{figure=resist_b1_5_a2_nu1_4.eps, width=6.5 cm,clip}
\end{center}
\vskip0.1cm
\caption{\label{fig:resist}
Resistivity computed from Eq.~(\ref{eq:logro}) for the localized phase of the
mean field model discussed in the text. 
We used  $A=2$ and $\nu = 1.4$ for the curves presented. The top figure shows the resistivity 
for parameters not very far from the localization phase transition ($B=1.5$),
while the bottom figure displays resistivity curves computed deep in the localized phase ($B=4$). 
A peak appears at $T_c$ due to the interplay of  magnetic ordering and localization, and is shifted to 
higher temperatures upon application of magnetic field.}
\end{figure}

Typical results are summarized in Fig.~\ref{fig:resist}
for $B=1.5$ and $B=4$. The resistivity curves are strikingly similar to the ones measured  
in various compounds in or in the vicinity of the localized phase, and clearly display a 
large peak at $T_C$ and a giant negative 
magneto-resistance \cite{manganites,Millis}.   This peak is simply a
consequence of reducing the localization length while entering the magnetic phase, and 
has nothing to do with critical fluctuations (which may also lead to 
additional contributions \cite{Felix}). 
The magnetic field dependence of the data also agrees qualitatively with the one 
 seen in the experiments: The peak is getting flat and shifts upward
with   increasing magnetic field. One of the most important properties of the 
experimental data is that the resistivity curves corresponding to different 
magnetization {\em do not cross}. The theory of Ref.~\cite{nagaev},
{\em e.g.} does not seem to satisfy this criterion \cite{gurin}, while in our theory, this is a natural 
consequence of the reduction of $\xi$. Note that in the absence of 
localization effects / disorder, 
the resistivity would {\em  not} display a peak at $T_c$ \cite{brey,furukawa}.
We find a similar peak structure in the metallic phase, however, there the precise shape of 
the anomaly  depends also on the
assumption made for the temperature dependence of the dephasing length, $L_\varphi(T)$.

In conclusion, in the present paper we have studied the interplay of disorder and magnetization 
in disordered local moment ferromagnets. We proposed a unified framework to study the 
localization  phase transition in these materials and argued that a unique  beta function 
can be used to describe the localization 
phase transition in these materials. We verified the above hypothesis for a simple 
model of disordered local moment ferromagnets. The scaling approach of this paper allowed us to 
estimate the temperature and magnetic field dependence of the 
resistivity in the localized phase of the mean field model studied. 
The obtained resistivity curves display a peak in the resistivity at the critical point, 
due to the interplay of magnetism and disorder. This resistivity maximum  is gradually 
suppressed and shifted towards higher temperatures upon application of a magnetic field, and the computed 
resistivity curves do not cross. Our simple theory thus seems to explain all basic features 
of the resistivity anomalies observed in many feromagnetic semiconductors in the 
'localized' phase and some of the manganites. 

We would like to thank J.K. Furdyna, P. Schiffer, I. Varga, T. Wojtowicz, and especially 
Peter Littlewood for valuable discussions. We would also like to thank P. Schiffer and B.-L. 
Sheu for sharing their results prior to publication. This research has been supported by
NSF-MTA-OTKA Grant No. INT-0130446, Hungarian Grants No. OTKA
T038162, T046267, and T046303, and the European 'Spintronics' RTN
HPRN-CT-2002-00302. 
B.J. was supported by NSF-NIRT award DMR02-10519 and by the Alfred P. Sloan Foundation.

\vspace{-0.8cm}

\end{document}